\documentclass[preprint,showpacs,prd,tightenlines,endfloats*]{revtex4-1}
\usepackage{graphicx}
\usepackage{amssymb}
\usepackage{amsmath}
\usepackage{bm}
\usepackage{natbib}
\newcommand{\p}[1]{(\ref{#1})}
\def\Journal#1#2#3#4{{#1} {\bf #2}, #3 (#4)}

\def\prd{{Phys. Rev.} D}

\def\apj{Astrophys. J.}

\def\epjc{Eur. Phys. J. C}

\begin{document}
\title[]{Comment on ``Covariant Tolman-Oppenheimer-Volkoff equations. II. The
anisotropic case'' 
}
\author{A. A.  \surname{Isayev}}
\email{isayev@kipt.kharkov.ua}
 \affiliation{Kharkov Institute of Physics and Technology, Academicheskaya
Street 1,
 Kharkov, 61108, Ukraine }
%

\date{\today}

\begin{abstract}
Recently, the covariant formulation of the Tolman-Oppenheimer-Volkoff (TOV)
equations for studying the equilibrium structure of a spherically symmetric
compact star in the presence of the pressure anisotropy in the interior of a
star was presented in Ref.~\cite{PRD18Carloni} [Phys. Rev. D \textbf{97} (2018) 124057]. It was suggested
there that the anisotropic solution of these equations can be obtained by
finding, first, the solution of the common TOV equations for the isotropic
pressure, and then by solving the differential equation for the anisotropic
pressure whose particular form was established on the basis of the covariant TOV
equations. It turns out that the anisotropic pressure determined according to
this scheme has a nonremovable singularity $\Pi\sim\frac{1}{r^2}$ in the center
of a star, and, hence, the corresponding anisotropic solution cannot represent a
physically relevant model of an anisotropic compact star. A new scheme for constructing the anisotropic solution, based on the covariant TOV equations, is suggested, which leads to the regularly behaved physical quantities in the interior of a star. A new algorithm is applied to build model anisotropic strange quark stars  with the MIT bag model equation of state.

\end{abstract}



\maketitle

The presence of the pressure anisotropy in the interior of a compact star should
be considered more like a common feature   than an exception.  The sources
of the pressure anisotropy can be very different such as the existence  of a
solid core, the relativistic nature of the nuclear interaction at high
densities, the presence of strong magnetic fields inside a star,~etc. The
equilibrium configuration of a spherically symmetric anisotropic compact star
can be studied on the basis of the  Tolman-Oppenheimer-Volkoff (TOV) equations,
generalized with account of the pressure anisotropy~\cite{ApJ74Bowers}.
Recently, these equations were presented in the covariant form with the help of
1+1+2 covariant formalism in the work~\cite{PRD18Carloni}. 
There are several strategies to find solutions of
these equations. One strategy is  to set the anisotropy parameter in the
specific preassigned form, which, together with the equation of state (EoS), can
be used to find  the energy density $\mu$, the
transverse $p_\perp$ and radial $p_r$ pressures, and the unknown metric
functions $A, B$ (in notations of Ref.~\cite{PRD18Carloni}). An example of such
an approach can be found, e.g., in Ref.~\cite{ApJ74Bowers}. The other strategy
is to somehow reduce the problem to the isotropic case, and then  to build the
anisotropic solution by properly modifying  the obtained isotropic one. Such an
approach was followed recently, e.g.,  in Refs.~\cite{EPJC16Shojai,PRD17I}.
Yet another strategy was suggested in Ref.~\cite{PRD18Carloni}, and is based on
the separation of the isotropic and anisotropic degrees of freedom in the
covariant TOV equations. After the separation, these equations take the form
(cf. Eqs.~(35) in Ref.~\cite{PRD18Carloni}):

\begin{gather}
 P_{,\rho}+P^2-P\Bigl(\mathbb{M}-3\mathcal{K}+\frac{7}{4}
\Bigr)\nonumber\\-\Bigl(\frac{1}{4}-\mathcal{K}\Bigr)\mathbb{M}=-\mathbb{P}_{,
\rho}-\mathbb{P}^2+\mathbb{P}\Bigl(-2P+\mathbb{M}-3\mathcal{K}+\frac{1}{4}\Bigr)
,\nonumber\\
 \mathcal{K}_{,\rho}=-2\mathcal{K}\Bigl(\mathcal{K}-\frac{1}{4}-\mathbb{M}\Bigr)
,\label{cov_TOV}
\end{gather}
where $P=B(\rho)p$ and $\mathbb{P}=B(\rho)\Pi$, $p\equiv \frac{p_r+2p_\perp}{3}$
and $\Pi\equiv \frac{2(p_r-p_\perp)}{3}$ being the isotropic and anisotropic 
pressure, respectively. The isotropic pressure terms are gathered in the
left-hand side (lhs) and the anisotropic pressure  terms are collected in the
right-hand side (rhs) of the first equation. Note the difference in sign in the
last term in the lhs, and in the first term and in the term with $P$ in the
brackets in the rhs of the first equation, compared to the corresponding
equation~(35) in Ref.~\cite{PRD18Carloni}. Eqs.~\p{cov_TOV} are nothing else
than the rewritten equations~(23) of Ref.~\cite{PRD18Carloni}, and the
difference in sign can be readily checked. Based on the structure of the
covariant TOV equations~\p{cov_TOV}, the following algorithm for constructing
the anisotropic solution was proposed in Ref.~\cite{PRD18Carloni}. First, to
find the solution of the common TOV equations, supplemented with the EoS
$\mathbb{M}=\mathbb{M}(P)$, for the isotropic pressure $P$ at vanishing
$\mathbb{P}$. In this way, there will be determined the coordinate dependence of
the functions $P, \mathcal{K}$ and $\mathbb{M}$. Then the obtained solution  of
the TOV equations in the isotropic case should be substituted into the equation
for the anisotropic pressure $\mathbb{P}$:
\begin{equation}
\mathbb{P}_{,\rho}+\mathbb{P}^2-\mathbb{P}\Bigl(-2P+\mathbb{M}-3\mathcal{K}
+\frac{1}{4}\Bigr)=0.\label{ani_P}
\end{equation} 

While the equation for $\mathcal{K}$  in this scheme is of the Bernoulli type
and  admits formal integration, the equation for the isotropic pressure
$P$, in view of the arbitrariness of the EoS $\mathbb{M}=\mathbb{M}(P)$, can be
integrated only numerically. For this reason, finding the anisotropic pressure
$\mathbb{P}$, in the general case, also requires  numerical integration. 

In order to check this algorithm for constructing the anisotropic solution, we will
 rewrite the covariant TOV equations in this scheme in the usual form with
the radial coordinate $r$ as an independent variable. 
The TOV equations in the isotropic case read (using the notations of
Ref.~\cite{PRD18Carloni} and the system of units with $c=1$):
\begin{align}
 p_{,r}&+G\frac{(\mu+p)(m(r)+4\pi pr^3)}{r(r-2Gm(r))}=0,\label{iso_TOV}\\
 m_{,r}&=4\pi\mu r^2.\label{m}
\end{align}
 Note that the differential equation for the metric function $B(r)$ was
rewritten in terms of the local mass function $m(r)$ which are related by
$B(r)=\Bigl(1-\frac{2Gm(r)}{r}\Bigr)^{-1}$. Eqs.~\p{iso_TOV} and \p{m},
supplemented with the EoS $\mu=\mu(p)$, should be solved together with the
initial conditions $p(0)=p_0, m(0)=0$, $p_0$~being the central isotropic
pressure. Eq.~\p{ani_P}, after straightforward transformations, takes the form
\begin{align}
 \Pi_{,r}+\frac{\Pi}{r}\Bigl(G\frac{m(r)+4\pi(\mu+2p+\Pi)r^3}{r-2Gm(r)}
+3\Bigr)=0.\label{Pi}
\end{align}
Note that solution of  Eqs.~\p{iso_TOV}--\p{Pi} should lead to the singularity--free
physical quantities in the interior of a star.  As follows from
Eq.~\p{Pi}, the gradient $\Pi_{,r}$ will be finite at $r=0$, if $\Pi\sim
r^\alpha$, $\alpha\geqslant1$ at $r\rightarrow0$. Given this asymptotic behavior
and  taking into account that $m(r)\sim r^3$ at $r\rightarrow0$, in the leading
order approximation on small $r$ Eq.~\p{Pi} reads 
 $\Pi_{,r}+\frac{3\Pi}{r}=0$,
which can be fulfilled only if $\alpha=-3$, i.e., $\Pi\sim \frac{1}{r^3}$ at
$r\rightarrow0$, that contradicts to the constraint $\alpha\geqslant1$. If the
anisotropic pressure $\Pi$ has a singularity at the origin, then the term with
$\Pi$ in the numerator of the fraction in the brackets in the lhs of  Eq.~\p{Pi}
is of relevance as well, and, after retaining it,  in the leading order
approximation one gets
\begin{align}
 \Pi_{,r}+\frac{3\Pi}{r}+4\pi G\Pi^2 r=0.
\end{align}
Substituting to the last equation $\Pi=C r^\alpha$, one can see that it can be
satisfied at any small $r$, only if $\alpha=-2$,  and, simultaneously,
$C=\frac{2}{\pi G}$.  Therefore, the anisotropic pressure $\Pi$, determined
according to Eq.~\p{Pi}, has a nonremovable singularity $\Pi\sim \frac{1}{r^2}$
at $r=0$, and, 
 hence, the
corresponding anisotropic solution cannot represent  a physically relevant model
of an anisotropic compact star. Note that this conclusion was reached on the basis of a general asymptotic analysis of Eq.~\p{Pi} only under the assumption  that the energy density $\mu$ and the isotropic pressure $p$ are regular functions
at the center of a star. This analysis doesn't rely on any specific form of the functions $\mu$ and~$p$.

In order to get the regularly behaved anisotropic pressure at the center of a star, let us present the equation relating the isotropic $P$ and anisotropic $\mathbb{P}$ pressures in Eq.~\p{cov_TOV} in a different form by carrying the term $\bigl(\frac{1}{4}-\mathcal{K}\bigr)\mathbb{M}$ to the rhs:

\begin{gather}
 P_{,\rho}+P^2-P\Bigl(\mathbb{M}-3\mathcal{K}+\frac{7}{4}
\Bigr)\nonumber\\=-\mathbb{P}_{,
\rho}-\mathbb{P}^2+\mathbb{P}\Bigl(-2P+\mathbb{M}-3\mathcal{K}+\frac{1}{4}\Bigr)+\Bigl(\frac{1}{4}-\mathcal{K}\Bigr)\mathbb{M}.
\label{ccov_TOV}
\end{gather}

A new algorithm for constructing the anisotropic solution of the covariant TOV equations consists in the following. First, it is necessary to find the isotropic pressure from the differential equations 

\begin{gather}
 P_{,\rho}+P^2-P\Bigl(\mathbb{M}-3\mathcal{K}+\frac{7}{4}
\Bigr)=0,
\nonumber\\
 \mathcal{K}_{,\rho}=-2\mathcal{K}\Bigl(\mathcal{K}-\frac{1}{4}-\mathbb{M}\Bigr),
\label{c_iso_TOV}
\end{gather}
supplemented by the EoS $\mathbb{M}=\mathbb{M}(P)$. Obtained in this way functions $P, \mathcal{K}$ and $\mathbb{M}$ should be substituted in the differential equation for the anisotropic pressure

\begin{equation}
\mathbb{P}_{,\rho}+\mathbb{P}^2-\mathbb{P}\Bigl(-2P+\mathbb{M}-3\mathcal{K}
+\frac{1}{4}\Bigr)-\Bigl(\frac{1}{4}-\mathcal{K}\Bigr)\mathbb{M}=0.
\label{c_ani_P}
\end{equation} 

Note that in this new setup 
Eqs.~\p{c_iso_TOV} for finding the isotropic pressure are different from the common TOV equations in the isotropic case. Precisely, the equation for the isotropic pressure with the radial coordinate as an independent variable reads (cf. Eq.~\p{iso_TOV} for the common TOV equations)

\begin{equation}
p_{,r}+G\frac{(\mu+p)(m(r)+4\pi pr^3)}{r(r-2Gm(r))}-\frac{G m(r)\mu}{r(r-2Gm(r))}=0,
\label{n_iso_TOV}
\end{equation} 
while the second equation in \p{c_iso_TOV} goes over to Eq.~\p{m}.
It is also interesting to notice that Eq.~\p{c_ani_P} has no trivial solution with $\mathbb{P}\equiv0$, and, hence the pressure anisotropy represents an essential feature of such class of compact stars. 

After straightforward transformations, Eq.~\p{c_ani_P} for the anisotropic pressure with the radial coordinate as an independent variable takes the form
 
\begin{align}
 \Pi_{,r}+\frac{\Pi}{r}\Bigl(G\frac{m(r)+4\pi(\mu+2p+\Pi)r^3}{r-2Gm(r)}
+3\Bigr)+\frac{G m(r)\mu}{r(r-2Gm(r))}=0.\label{cPi}
\end{align}

Now, assuming that $\Pi=C^{\,\prime}
r^\alpha$, $\alpha\geqslant1$ at $r\rightarrow0$, in the leading order approximation one gets

\begin{align}\Pi_{,r}+\frac{3\Pi}{r}+\frac{4\pi G\mu_0^2r}{3}=0, \label{asi_Pi}
 \end{align}
where $\mu_0\equiv\mu(0)$ is the energy density at the center of a star. Eq.~\p{asi_Pi} can be satisfied at small $r$, only if $\alpha=2$ and 
$C^{\,\prime}=-\frac{4\pi G\mu_0^2}{15}$, i.e., $\Pi\sim
r^2$ at $r\rightarrow0$. Therefore, the anisotropic pressure, determined according to Eq.~\p{cPi} in a new scheme, is the regularly behaved function at the center of a star.

\begin{table}[tb]
\caption{The mass $M$ of an anisotropic strange quark star (in solar mass units), determined according to Eqs.~\p{m}, \p{n_iso_TOV} and  \p{cPi}, for different values of the central isotropic pressure $p_0$  within the MIT bag model at  $B=57$ MeV/fm$^3$.}\vspace{6mm}

 \hfill\hbox{
\begin{tabular}{c|c}

\hline 
$p_0$, $\frac{\textrm{MeV}}{\textrm{fm}^3}$ & $M/M_\odot$ \\
\hline 
0.001 & 6.7338 \\
\hline
0.01 & 6.7337 \\
\hline
0.1 & 6.7321 \\
\hline
1 & 6.7163 \\
\hline
10 & 6.5722 \\
\hline
20 & 6.4362 \\
\hline
30 & 6.3199 \\
\hline
40 & 6.2190 \\
\hline
50 & 6.1305 \\
\hline
60 & 6.0523 \\
\hline
70 & 5.9827 \\
\hline
80 & 5.9203 \\
\hline
90 & 5.8640 \\
\hline
100 & 5.8131 \\
\hline
\end{tabular}}\hfill
 \label{table1}
\end{table} 

In order to test this new algorithm, let us consider anisotropic strange quark stars within the MIT bag model with the massless quarks and the EoS  $\mu=3p+4B$, $B$ being the bag constant. The radius of a spherically symmetric anisotropic star is determined from the condition $p_r(R)=p(R)+\Pi(R)=0$, where isotropic $p(r)$ and anisotropic $\Pi(r)$ pressures, together  with the mass function $m(r)$, are obtained by solving the differential equations \p{m}, \p{n_iso_TOV}, \p{cPi} with the initial conditions $p(0)=p_0, m(0)=0, \Pi(0)=0$. The total mass of a compact star is found as $M=m(R)$. Table 1 presents the results of the numerical determination of the total mass of an anisotropic strange quark star in dependence on the central isotropic pressure $p_0$ at $B=57$ MeV/fm$^3$. It is seen that even small central pressures $p_0<1$~MeV/fm$^3$ produce heavy strange quark stars with $M>6M_\odot$ ($M_\odot$ being the solar mass). This is because the isotropic pressure, determined according to Eq.~\p{n_iso_TOV},  decreases considerably slower with the radial coordinate $r$ compared to that determined from the common TOV equation \p{iso_TOV}. Another peculiarity is that the total mass $M$ decreases with the central pressure $p_0$ for the whole range of the central pressures under consideration,  
contrary to the stability constraint $\frac{dM}{dp_0}>0$. This means that the obtained model massive anisotropic  strange quark stars are unstable with respect to radial oscillations. 

To summarize, the proposed algorithm in Ref.~\cite{PRD18Carloni} for
constructing the anisotropic solution of the TOV equations, based on solving the
common TOV equations for the isotropic pressure and Eq.~\p{ani_P} for the
anisotropic pressure (in the covariant formulation), leads to the singularly
behaved anisotropic pressure at the center of a star, and, hence, the
corresponding anisotropic solution cannot represent  a physically relevant model
of an anisotropic compact star. In this notice, it has been suggested a new scheme for constructing the anisotropic solution, based on the covariant TOV equations, which leads to the regularly behaved anisotropic pressure in the interior of a star.  This algorithm has been tested with respect to anisotropic strange quark stars within the MIT bag model. It turns out that the new algorithm gives rise to massive anisotropic strange quark stars with $M\gtrsim5M_\odot$, which are unstable with respect to radial oscillations. 

In the end, it would be also correctly to note that
Ref.~\cite{PRD18Carloni} contains a number of other suggestions on constructing
solutions of the TOV equations in the anisotropic case, mainly, in the
analytical form, whose discussion is, however,  beyond the scope of this
comment.


\end{document}